%% file: paper_joint.tex
\documentclass[10pt,floatfix,amsmath,amssymb,twocolumn,superscriptaddress,groupedaddress,nofootinbib,aps,prd]{revtex4-1}

\usepackage{nicefrac}
\usepackage{braket}
\usepackage{float}
\DeclareMathOperator{\tr}{tr}
\newcommand{\SEE}{
	S_{\text{EE}}
}
\newcommand{\eqnPunct}{
}

\usepackage{tikz}
\usepackage{pgfplots}
\pgfplotsset{compat=1.14}
\usetikzlibrary{calc}
\newcommand\tikzClose[2]{
	\draw[rounded corners = 2pt] #1
		--++ (0.2,0.2)
		--++ (0.2,-0.2)
		-- ($#2 + (0.4,0)$)
		--++ (-0.2,-0.2)
		--++ (-0.2,0.2)
		-- #2;
	
}

\usepackage{placeins} 
\usepackage{diagbox}  
\usepackage{enumitem} 
\usepackage{graphicx}

\newcommand{\eref}[1]{%
	(\ref{#1})%
}

\newcommand{\comment}[1]{}

\definecolor{fakultaet}{RGB}{0,137,147}  

\usepackage[normalem]{ulem}

\def\AThree{ 0.030 \pm 0.003}
\def\BThree{ -0.328 \pm 0.112}
\def\CThree{ 1.422 \pm 0.068}
\def\SlopeThree{-0.030\pm 0.013}

\def\AFour{ 0.029 \pm 0.001}
\def\BFour{ -0.132 \pm 0.044}
\def\CFour{ 0.997 \pm 0.029}
\def\SlopeFour{-0.070\pm 0.025}

\def\lmaxTwo{1.1}
\def\lmaxThree{1.3}
\def\lmaxFour{0.7}
\begin{document}
    \sloppy
    \title{Lattice study of R\'enyi entanglement entropy in $SU(N_c)$ lattice Yang-Mills theory with $N_c = 2, 3, 4$}
    \author{Andreas~Rabenstein}
    \email{andreas.rabenstein@physik.uni-r.de}
    \affiliation{
    	Institute for Theoretical Physics, University of Regensburg,
    	93040 Regensburg, Germany
    }
    \author{Norbert~Bodendorfer}
    \email{norbert.bodendorfer@physik.uni-r.de}
    \affiliation{
    	Institute for Theoretical Physics, University of Regensburg,
    	93040 Regensburg, Germany
    }
    \author{Pavel~Buividovich}
    \email{pavel.buividovich@physik.uni-r.de}
    \affiliation{
    	Institute for Theoretical Physics, University of Regensburg,
    	93040 Regensburg, Germany
    }
    \affiliation{
        Institut f\"ur Theoretische Physik, Justus-Liebig-Universit\"at,
        35392 Giessen, Germany}
    \author{Andreas~Sch\"afer}
    \email{andreas.schaefer@physik.uni-r.de}
    \affiliation{
    	Institute for Theoretical Physics, University of Regensburg,
    	93040 Regensburg, Germany
    }

    \begin{abstract}
    
    	We consider the second R\'enyi entropy $S^{(2)}$ in pure lattice gauge theory with $SU(2)$, $SU(3)$ and $SU(4)$ gauge groups, which serves as a first approximation for the entanglement entropy and the entropic $C$-function. We compare the results for different gauge groups using scale setting via the string tension. We confirm that at small distances $l$ our approximation for the entropic $C$-function $C(l)$, calculated for the slab-shaped entangled region of width $l$, scales as $N_c^2 - 1$ in accordance with its interpretation in terms of free gluons. At larger distances $l$ $C(l)$ is found to approach zero for $N_c = 3, 4$, somewhat more rapidly for $N_c = 4$ than for $N_c = 3$. This finding supports the conjectured discontinuity of the entropic $C$-function in the large-$N$ limit, which was found in the context of AdS/CFT correspondence and which can be interpreted as transition between colorful quarks and gluons at small distances and colorless confined states at long distances. On the other hand, for $SU(2)$ gauge group the long-distance behavior of the entropic $C$-function is inconclusive so far. There exists a small region of lattice spacings yielding results consistent with $N_c=3,4$, while results from other lattice spacings deviate without clear systematics. We discuss several possible causes for discrepancies between our results and the behavior of entanglement entropy in holographic models.
    \end{abstract}

	\maketitle

	
	\section{Introduction}

Entanglement entropy has become a heavily studied field of research in recent years. For example, it is widely used in quantum information theory, where many other entanglement measurements, as e.g. mutual entropy, exist \cite{Cerf:1996nb}, as well as in connection with the gauge/gravity duality \cite{Ryu:2006bv}. It can also be used as a universal order parameter of quantum phase transitions, as it is done, e.g. for $2+1$ dimensional topological field theories, where one relates entanglement entropy to the quantum dimension \cite{Kitaev:2005dm}.

More generally, in the context of quantum field theory, entanglement entropy can be considered as a counter of the effective number of degrees of freedom, which is related to the central charge in two-dimensional conformal field theories (CFTs) \cite{Casini:hep-th/0405111}. Much like the central charge, entanglement entropy changes monotonically under renormalization group (RG) flow, and obeys a generalization of Zamolodchikov's $c$-theorem \cite{ZamolodchikovJETPLett1986}. In higher-dimensional field theories, entanglement entropy can be related to Cardy's $A$-function which generalizes Zamolodchikov's $c$-function \cite{Myers:1011.5819} and which is also monotonous under RG flow \cite{CardyPhysLettB1988,Komargodski:1107.3987,Mezei:1202.2070}.

In strongly coupled quantum field theories, the entanglement entropy turns out to be notoriously hard to compute from first principles either numerically or analytically unless the models are sufficiently symmetric or low-dimensional, see e.g. \cite{Casini:2009sr} for an overview. In the light of this difficulty, the AdS/CFT duality offers an attractive alternative route to calculate it in terms of the area of minimal surfaces in a dual gravitational theory \cite{Ryu:2006bv}. While this computation is a priori justified only for large number of colors $N_c$ and strong coupling in $\mathcal N=4$ super Yang-Mills theory, it may still provide valuable guidance to understand the generic behavior of entanglement entropy in strongly coupled gauge theories. Indeed, the computation provided in \cite{Klebanov:2007ws} using a confining background predicted an abrupt transition at some $l = l_c$ in the dependence of entanglement entropy on the characteristic size $l$ of the entangled region, at which the scaling of the entanglement entropy changes from $O(N_c^2)$ to $O(1)$. Since the entanglement entropy is expected to count an effective number of degrees of freedom at energy scale $1/l$ \cite{Mezei:1202.2070}, the transition from $O(N_c^2)$ to $O(1)$ scaling of the entropy can be interpreted as a transition between colorful quarks and gluons at small distances and colorless hadrons and glueballs at long distances. While such scaling laws at asymptotically short and large distances can be readily understood in terms of asymptotic freedom and confinement, the discontinuous nature of the transition is a non-trivial prediction for theories with a Hagedorn-type spectrum, which also include holographic models.

This non-trivial prediction was taken as a motivation to analyze the spatial dependence of the entanglement entropy in pure $SU(2)$ \cite{Buividovich:2008kq} and $SU(3)$ \cite{Itou:2015cyu} Yang-Mills theory, approximating it with a R\'enyi entanglement entropy for the sake of numerical convenience. Somewhat surprisingly, there is a qualitative difference in the results: while \cite{Buividovich:2008kq} seems to show an abrupt transition after an initial enhancement, \cite{Itou:2015cyu} is compatible with both an abrupt and a smooth transition within statistical errors, but leans more towards a smooth transition. 

In this work, we continue this line of research and present a comparative high-statistics study of the R\'enyi entanglement entropy in lattice gauge theories with $SU(2)$, $SU(3)$ and $SU(4)$ gauge groups. Our aims are to study the scaling of entanglement entropy with $N_c$ and to check whether the dependence of the entanglement entropy on the size of the entangling region approaches a discontinuous function at large $N_c$. Next to better understanding $SU(N_c)$ Yang-Mills theory, this is of obvious interest to quantify differences between holographic and real world gauge theories. 

We work with pure $SU(N_c)$ Yang-Mills theory, as already computing the second R\'enyi entropy to approximate the entanglement entropy turns out to be computationally expensive. Owing to this, we are also not able to exceed a lattice size of $16^3 \times 32$ or perform a continuum extrapolation\footnote{Investigations of the impact of finite volume effects, continuum extrapolation, and replica number dependence for $SU(3)$ showed them to be smaller than the statistical error \cite{Itou:2015cyu}.}. 

We are able to clearly establish the scaling of R\'enyi entanglement entropy with the number of gluon states $N_c^2 - 1$ at short distances. While the transition between $O(N_c^2)$ and $O(1)$ scalings of the entropy at short and large distances appears to be smooth for our data for all $N_c$ which we consider (as it should be for a finite-$N_c$ gauge theory on a finite lattice), we find indications that for $N_c = 4$ the entanglement entropy changes faster in the transition region than for $N_c = 3$. This supports the scenario of a discontinuous transition in the large-$N_c$ limit, in agreement with the arguments of \cite{Klebanov:2007ws} which predict a discontinuous behavior of entanglement entropy for any field theory with a Hagedorn-type spectrum.

This paper is structured as follows: we start by reminding the reader of basic definitions and properties of entanglement entropy, also from a holographic perspective, in section \ref{sec:EE}.
In section \ref{sec:Measurement}, we will recall how one can measure entanglement entropy on the lattice, approximating it with the second R\'enyi entanglement entropy. Section \ref{sec:Results} contains our numerical results, their interpretation, and a comparison between different $N_c$. Finally we will conclude and point out possible future directions in section \ref{sec:Conclusion}.

\section{Entanglement and R\'enyi entropy} \label{sec:EE}

Intuitively, entanglement entropy measures the quantum correlations of quantum fields in two complementary spatial regions $\mathcal{A}$ and $\bar{\mathcal{A}}$. Hence, we start with a bipartition $\Sigma_t = \mathcal{A} \cup \bar{\mathcal{A}}$ of a fixed time slice $\Sigma_t$ as shown in the left picture of figure~\ref{fig:Bipartition}.
Next we write the Hilbert space $\mathcal{H}$ as a product space of the bipartition, i.e. $\mathcal{H} = \mathcal{H}_\mathcal{A} \otimes \mathcal{H}_{\bar{\mathcal{A}}}$. For non-gauge theories this factorization is unique, but for gauge theories there are ambiguities how to factorize the Hilbert space. In this work we use the so-called maximally gauge invariant way~\cite{Aoki:2015bsa,Buividovich:0806.3376}.
In this method, one cuts along the gauge links and then decides to which region, $\mathcal{A}$ or $\bar{\mathcal{A}}$, the link degree of freedom belongs to. 
This  is visualized in the right picture of figure~\ref{fig:Bipartition}.
\input{tikzBipartition}
To measure the entanglement between the two regions, we take a partial trace over region $\mathcal{A}$ and define the reduced density matrix to be
\begin{align}
	\hat{\rho}_{\mathcal{A}} = \tr_{\bar{\mathcal{A}}}\left(\hat{\rho}\right) 
\end{align}
where $\hat{\rho} = \ket{\psi}\bra{\psi}$ is the density matrix of the whole system.
The entanglement entropy is defined as the von Neumann entropy of the reduced system, i.e.
\begin{align}
	\SEE = - \tr_{\mathcal{A}}\left(
			\hat{\rho}_{\mathcal{A}} \log\left(\hat{\rho}_\mathcal{A}\right)
		\right)
	\eqnPunct .
\end{align}
This definition can be generalized to R\'enyi entropies which are given by~\cite{Renyi:1961aaa,Renyi:1965aaa}
\begin{align}\label{eqn:DefRenyi}
	S^{(q)} = \frac{1}{1-q}\log\left(
		\tr_{\mathcal{A}}
		\left(
			\hat{\rho}_{\mathcal{A}}^q
		\right)
	\right)
	\eqnPunct \forall\ q \in \mathbb{N}, ~ q \geq 2  \eqnPunct .
\end{align}
For most systems, as also for the system we will consider, this definition can be analytically continued to $q\in\mathbb{R}_+$.
With this, one can show that the limit $q\rightarrow 1 $ of the R\'enyi entropies gives the entanglement entropy, i.e.
\begin{align}\label{eqn:RelationRenyiEntanglement}
\SEE= 	\lim_{q\rightarrow 1 }S^{(q)} 
	\eqnPunct .
\end{align}
We emphasize that computing $\SEE$ is hard in Monte-Carlo simulations due to the logarithm of the reduced density matrix, whereas $S^{(q)}$ for $q \in \mathbb N, ~ q \geq 2$ is relatively simple to evaluate. On the lattice, it seems most economical to use $S^{(2)}$ as an approximation for $\SEE$, which can be motivated, in particular, by looking at the $q$-dependence of $S^{(q)}$ in simplified models \cite{Calabrese:2004eu,Casini:cond-mat/0511014} such as a free massive scalar field. In this case the $l$-dependent terms in $S^{(q)}$ differ from corresponding terms in the entanglement entropy by a factor $\frac{q+1}{2q}$, see e.g. equation~(86) in \cite{Casini:cond-mat/0511014}. For the second R\'enyi entropy with $q=2$ this gives a factor of $3/4$. Higher R\'enyi entropies differ from the entanglement entropy even more. On the other hand, the relative difference between $S^{(2)}$ and $S^{(3)}$ appears to be less than $10\%$, which in practice is comparable with the numerical accuracy of measuring entanglement entropy. These considerations make it apparent that computing higher R\'enyi entropies in order to improve the extrapolation to $q=1$ would be too costly due to the very mild dependence of $S^{(q)}$ on $q$ for larger $q$. We therefore expect a systematic error of at least $25\%$ due to using the second R\'enyi entropy instead of the entanglement entropy. However, explicit calculations in free field theories suggest that the functional dependence on $l$ is the same in both cases \cite{Calabrese:2004eu,Casini:cond-mat/0511014}.

It should be noted that the alternative formula 
\[
	\SEE = - \lim_{q \rightarrow 1} \frac{\partial}{\partial q} \log \text{tr} \left( \hat{\rho}_{\mathcal{A}}^q \right)
\]
leads to the same result once the derivative is approximated as $f'(q=1)=f(q=2)-f(q=1)$ on the lattice, i.e. one also measures precisely $S^{(2)}$.

The entanglement and R\'enyi entropies typically contain non-universal UV divergent terms, which have to be carefully removed in order to access the universal low-energy properties of the theory. From a theoretical viewpoint, extraction of universal UV-finite terms from the entanglement entropy has been most clearly worked out for finite and scalable entangled regions, such as the interior of a sphere \cite{Mezei:1202.2070}. For such regions, characterized by a single dimensionful size parameter $l$, the universal UV-finite contribution $C(l)$ to the entanglement entropy can be extracted as 
\begin{align}
\label{Mezei_Cfunc}
 C(l) = l \partial_l (l \partial_l S_E(l) - 2 S_E(l)) .
\end{align}

However, for a realistic lattice gauge theory simulation on a square lattice it is practically impossible to implement a smooth entangling surface such as a sphere. Smooth surfaces could only be approximated as collections of polygons, which would lead to a plethora of additional contributions due to surface edges \cite{Myers:1505.07842}. In what follows we will hence restrict ourselves to a slab-shaped entangled region of width $l$ embedded in a three-dimensional square lattice of size $L$ with periodic boundary conditions, a geometry which is typically used in lattice gauge theory simulations \cite{Buividovich:2008kq,Itou:2015cyu} and which was also considered in detail in theoretical works \cite{Klebanov:2007ws,Ryu:hep-th/0605073,Casini:cond-mat/0511014}. Due to the periodic boundary conditions in space directions, the corresponding entangling surface consists of two parallel planes and thus has neither extrinsic curvature nor sharp corners.

Assuming that the lattice size $L$ is reasonably big, for such a slab-shaped entangled region it is natural to consider the entanglement entropy per unit area of the slab surface. Dimensional analysis and semi-analytic estimates \cite{Ryu:hep-th/0605073,Casini:cond-mat/0511014,Klebanov:2007ws} suggest that at small slab width $l$ the entanglement entropy per unit area in $d=3+1$ dimensions behaves as
\begin{align}
\label{SEE_scaling}
	\SEE/L^2 = \frac{A'}{a^2} - \frac{C}{l^2} + (finite) ,
\end{align}
where $a$ is the lattice spacing which sets the UV cutoff scale $\Lambda_{UV} \equiv a^{-1}$ and $(finite)$ denotes UV-finite terms which do not diverge as $a \rightarrow 0$ or $l \rightarrow 0$. For quantum field theories with a mass gap, at large $l$ $\SEE$ should approach a constant value.

The coefficient $C$ in (\ref{SEE_scaling}) is believed to be universal and similar to the $A$-function (in the terminology of \cite{CardyPhysLettB1988,Myers:1011.5819,Komargodski:1107.3987}), a higher-dimensional generalization of central charge of 2D conformal field theories which serves as a counter of the number of effective degrees of freedom in a theory. In particular, it is expected to decrease monotonically along the renormalization group flow, similarly to Zamolodchikov's $c$-function.

To extract $C$ from the entanglement entropy (\ref{SEE_scaling}), one can apply the differential operator $l^3 \partial_l$ to the entanglement entropy (\ref{SEE_scaling}): 
\begin{align}
\label{eq:EntropicCFunctionDef}
	C(l) = \frac{l^3}{2 \, L^2} \frac{\partial\SEE(l)}{\partial l} .
\end{align}
At small $l \sim a$ $C(l)$ counts the number of degrees of freedom at the UV cutoff scale (such as e.g. asymptotically free gluons in QCD). However, at intermediate values of $l$, when finite terms in (\ref{SEE_scaling}) become important, the coefficient $C(l)$ defined by (\ref{eq:EntropicCFunctionDef}) acquires a nontrivial $l$-dependence and can be interpreted as a counter of the number of degrees of freedom at scale $l$. By analogy with higher-dimensional generalizations of Zamolodchikov's $c$-theorem \cite{CardyPhysLettB1988,Myers:1011.5819,Komargodski:1107.3987} one expects that $C(l)$ would monotonically decrease with $l$.

For slab-shaped entangled regions such an interpretation has only been confirmed by explicit calculations in free field theories \cite{Ryu:hep-th/0605073,Casini:cond-mat/0511014} and in holographic models \cite{Klebanov:2007ws}. In this paper we follow \cite{Ryu:hep-th/0605073,Casini:cond-mat/0511014,Klebanov:2007ws} and consider $C$ as being approximately proportional to the effective number of degrees of freedom, referring to $C(l)$ as an entropic $C$-function \cite{Buividovich:2008kq,Nishioka:2009un,Itou:2015cyu}. We will see that our numerical results support such an interpretation.

A qualitative analysis of the entanglement entropy for a slab-shaped region (see e.g. equation~(4.5) in \cite{Ryu:hep-th/0605073}) suggests that the term $- \frac{C}{l^2}$ in (\ref{SEE_scaling}) might also contain an additional logarithmic factor: 
\begin{align}
\label{SEE_scaling_log}
	\SEE/L^2
	= 
	\frac{A'}{a^2}
	-
	\frac{C}{l^2}
	-
	\frac{C'}{l^2}\ln(l)
	+
	(finite) .
\end{align}
In practice, however, it is difficult to extract a relatively small $\ln(l)$ correction to the $l^{-2}$ scaling law. While one can construct a formal expression $C'(l) = l \frac{\partial}{\partial l} \left( \frac{l^3}{2 L^2} \frac{\partial\SEE(l)}{\partial l} \right)$ similar to (\ref{eq:EntropicCFunctionDef}) and (\ref{Mezei_Cfunc}) yielding the coefficient $C'$ in (\ref{SEE_scaling_log}), it yields a very noisy signal upon the replacement of a second derivative by finite differences, and is therefore not practical. We thus neglect possible $\ln(l)$ correction and use the expression (\ref{eq:EntropicCFunctionDef}) to extract the entropic $C$-function numerically.

Dimensional arguments suggest that at small slab width $l$ the R\'enyi entropies (and in particular the second R\'enyi entropy, which we actually measure) should also behave similarly to (\ref{SEE_scaling}), however, with coefficient $C$ which is in general different from the one for the entanglement entropy. As discussed above, for free field theory the relative error in $C$ is as large as $25\%$ \cite{Casini:cond-mat/0511014}.

Let us note that a wealth of results exist for the scaling behaviour of R\'enyi and entanglement entropies in lower, in particular two-dimensional, gapless field theories. The general lesson to be drawn from these studies is to be careful with the naive $C$-function definition \eqref{SEE_scaling} and the usage of the second R\'enyi entropy as an approximation for the entanglement entropy. As mentioned before, the coefficient $C$ may vary between the entanglement entropy and the second R\'enyi entropy in explicit examples (e.g. \cite{GliozziEntanglementEntropyAnd, TagliacozzoSimulationOfTwoDimensional}) and logarithmically diverging terms may occur (e.g. \cite{HelmesRenyiEntropyPerspective}). Additionally, so-called ``unusual corrections'' were found \cite{CardyUnusualCorrectionsTo, FagottiEntanglementEntropyOf}, including oscillations in $l$. The necessity to go to large lattices sizes and large $l$ to obtain a good match with CFT results was pointed out in \cite{AlbaEntanglementEntropyOf}. In general, accurate results can only be expected when all of the above corrections are considered \cite{SahooUnusualCorrectionsTo}. With this in mind, let us note that gapless lower-dimensional systems are very special and may or may not reflect the physics of four-dimensional Yang-Mills theory with a finite mass gap that we are considering in this paper. Due to a lack of analytic results concerning the entanglement and R\'enyi entropies in this case, we are forced to simply test the prescription \eqref{eq:EntropicCFunctionDef} against our numerical data.

In holographic models entanglement entropy can be calculated using the Ryu-Takayanagi formula~\cite{Ryu:2006bv} which states that entanglement entropy is proportional to the area of the minimal surface whose boundary coincides with the entangling region $\partial \mathcal{A}$, i.e.
\begin{align}
	\SEE^{\text{hol}} = \frac{A_\text{min}}{4 G_N \hbar}
	\eqnPunct
	~~~\text{   for $N_c\rightarrow\infty$ and $\lambda$ large.}
\end{align}
This formula was applied to confining backgrounds in \cite{Klebanov:2007ws} to find that there exists always a connected and a disconnected solution to the equations of motion which one has to consider to find the minimal area. The connected solution depends on the length $l$ of the subsystem while the disconnected one is constant for all $l$. Since these two solutions do not intersect smoothly we can conclude that the entropic $C$-function has a jump at $l = l_c $. At $l<l_c$ $C(l)$ scales as $N_c^2$ (to the leading order of $1/N_c$ expansion), while at $l > l_c$ $C(l)$ is of order of unity, as could be expected due to confinement at large distances.

In the next section we discuss how to calculate the entropic $C$-function on the lattice.

\section{Measuring R\'enyi entropy on the lattice} \label{sec:Measurement}
In this section we will describe how we can measure R\'enyi entropy on the lattice.
This has already been done for $SU(2)$ and $SU(3)$ pure Yang-Mills theory in~\cite{Buividovich:2008kq} and~\cite{Itou:2015cyu} respectively.
Therefore we will only briefly review the method and refer the interested reader to these publications.

From equation~\eref{eqn:DefRenyi} we see that we have to calculate the trace of powers of the reduced density matrix, i.e. $\tr_\mathcal{A}\left(\hat{\rho}_\mathcal{A}^q\right)$.
A strategy to solve this problem using path integrals was first proposed in~\cite{Calabrese:2004eu}, which we will review in a graphical way adapted to the lattice.
We start with the calculation of matrix elements of the density matrix $\hat{\rho}$ which we can write as
\begin{align}
	\bra{\psi_1}\hat{\rho}\ket{\psi_2}
	=
	\frac{1}{Z}
	\bra{\psi_1}e^{-\beta \hat{H}}\ket{\psi_2}
	=
	\begin{tikzpicture}[scale=0.7,baseline={([yshift=-.6ex]current bounding box.center)}]
		\input{tikz_density}
	\end{tikzpicture}
	\eqnPunct .
\end{align}

In this work, we always refer to the density matrix of the groundstate, i.e. $\hat{\rho} = \ket{0}\bra{0}$, which is obtained in the limit $\beta\rightarrow \infty$.
We have to understand the picture in this formula such that we integrate over all field configurations from Euclidean time $\tau_E = -\infty$ to $\tau_E = \infty$ with $\ket{\psi_{\nicefrac{1}{2}}}$ as boundary conditions.
On the lattice, this integral is discretized and then approximated using Monte Carlo techniques.
With this in mind, we can calculate the reduced density matrix by tracing over region $\bar{\mathcal{A}}$, i.e.
\begin{align}\nonumber
	\bra{\psi_1}\hat{\rho}_{\mathcal{A}}\ket{\psi_2}
	=
	\sum\limits_{\ket{\psi}}&
	\begin{tikzpicture}[scale=0.7,baseline={([yshift=-.6ex]current bounding box.center)}]
		\input{tikz_density_reduced}
	\end{tikzpicture}\\
	&=\hspace{1em}
	\begin{tikzpicture}[scale=0.7,baseline={([yshift=-.6ex]current bounding box.center)}]
		\input{tikz_density_reduced_deformed}
	\end{tikzpicture}
	\eqnPunct .
\end{align}
In the last step, we deformed the lattice such that we automatically get rid of the sum by identifying corresponding states $\ket{\psi}$.
It is important to note that the same number of degrees of freedom label a regular lattice and a deformed lattice. In particular, the diagonal link in the deformed lattice carries the same group element as the middle lower link. This feature will later allow to compute the derivative in the holographic $C$-function by comparing different cut lengths $l$ obtained from the same underlying lattice configurations.

Powers of the reduced density matrix can be computed by glueing copies of the above lattices together, 
i.e. 
\begin{align}
	\tr_{\mathcal{A}}\left(\hat{\rho}_{\mathcal{A}}^2\right)
	=\hspace{0.5em}
	\begin{tikzpicture}[scale=0.7,baseline={([yshift=-.6ex]current bounding box.center)}]
		\input{tikz_tr_density_reduced_squared}
	\end{tikzpicture}
	\ \ \ .
\end{align}
This procedure easily generalizes to arbitrary integer powers as
\begin{align}\label{eqn:trRedSquaredLattice}
	\tr_{\mathcal{A}}\left(\hat{\rho}_{\mathcal{A}}^q\right)
	=
	\frac{Z[\mathcal{A},q,T]}{Z^q[T]}
\end{align}
where $Z[T]$ is the partition function of a normal lattice and the partition function $Z[\mathcal{A},q,T]$ refers to a lattice which has a temporal period of $q\times T$ in $\mathcal{A}$ and $q$ subsystems with period $T$ in $\bar{\mathcal{A}}$.
Figure \ref{fig:LatticePeriodicity} provides an alternative graphical depiction already used in \cite{Itou:2015cyu}.
\begin{figure}
	\centering
	\begin{tikzpicture}[scale=0.6]
		\input{tikz_periodicity}
	\end{tikzpicture}
	\caption{
		Visualization of the deformed geometry as it was already shown in~\cite{Itou:2015cyu}.
		The periodicity in region $\mathcal{A}$ (right, red) is $2T$ while in $\bar{\mathcal{A}}$ (left, green) we have two copies with periodicity $T$. 
		The crosses, squares and triangles indicate the identifications which have to be made.
}
	\label{fig:LatticePeriodicity}
\end{figure}

Using~\eref{eqn:DefRenyi} and~\eref{eqn:trRedSquaredLattice} we can write the R\'enyi entropies as
\begin{align}\label{eqn:RenyiFreeEnergy}
	S^{(q)} 
	= 
	\frac{1}{q-1}F[\mathcal{A},q,T] - \frac{q}{q-1}F[T]
\end{align}
where we used the relation $F = -\log Z$ between the free energy $F$ and the partition function $Z$. As already mentioned, we will approximate the entanglement entropy by the second R\'enyi entropy $S^{(2)}$. To get an approximation for the entropic $C$-function, we have to calculate the derivative of $S^{(2)}$ with respect to $l$. We do this by using the central finite difference approximation, i.e. 
\begin{align}\label{eq:ApproximateDerivative}
	\frac{\partial}{\partial l} S^{(2)}(l-\nicefrac{a}{2})
	\approx
	\frac{F[l-a,2,T] - F[l,2,T]}{a}
	\eqnPunct .
\end{align}
The term $\sim F[T]$ in~\eref{eqn:RenyiFreeEnergy} vanishes since it does not depend on $l$.

So far, we have reduced the problem of calculating the entropic $C$-function to measuring differences of free energies on the deformed lattice geometry defined above.

Such differences can be calculated using the relation between the free energy and the partition function and the Fundamental Theorem of Calculus~\cite{Fodor:2007sy,Endrodi:2007tq}. Namely, the difference between the free energies $F_2 = - \log Z_2$ and $F_1 = - \log Z_1$ which correspond to the lattice actions $S_2[U]$ and $S_1[U]$ can be represented as an integral of the form
\begin{align}
	F_2 - F_1 
	= 
	-\log Z_2 + \log Z_1
	=
	-\int_0^1d\alpha \frac{\partial}{\partial\alpha} \log Z(\alpha)
\end{align}
where $Z(\alpha)$ is the partition function which interpolates in some way between $Z_2$ for $\alpha=1$ and $Z_1$ for $\alpha=0$. The simplest partition function satisfying this requirement can be constructed by linearly interpolating between the actions $S_2[U]$ and $S_1[U]$:
$
	Z(\alpha) = \int D[U]\exp
	\left(
		-(1-\alpha) S_1[U]	- \alpha S_2[U]
	\right)
$
where $\int D[U]$ refers to the path integral over all gauge links $U$. Then we can calculate the derivative with respect to $\alpha$ to obtain
\begin{align}\label{eqn:DifferenceFreeEnergies}
	F_2 - F_1
	=
	\int_{0}^1 d\alpha
		\left<
			S_2[U] - S_1[U]
		\right>_\alpha
	\eqnPunct .
\end{align}
In this formula $\left<\cdot\right>_\alpha$ denotes the average with weight of the interpolating partition function defined above. Coming back to our earlier remark, we note that it was crucial in the last step that we can interpret the same underlying lattice data as different cut lengths. 

With this procedure, we can calculate the entropic $C$-function by the following steps:
\begin{enumerate}
	\item Generate gauge configurations with interpolating action
		\begin{align}
			S_\text{int} = (1-\alpha) S_l[U] + \alpha S_{l+1}[U]
		\end{align}
		on the deformed lattice (see figure~\ref{fig:LatticePeriodicity}).
		The subscripts $l$ and $l+1$ refer to the length of the cut.
	\item Measure $S_{l+1}-S_l$ on these configurations for several values of $\alpha$
	\item Integrate over $\alpha$ by interpolating with cubic splines to approximate the first derivative of the second R\'enyi entropy using \eqref{eq:ApproximateDerivative} and \eqref{eqn:DifferenceFreeEnergies}
	    \begin{align}
	        \frac{\partial S^{(2)}}{\partial l}\bigg\vert_{l-\nicefrac{a}{2}}
	        = \int_{0}^{1} d\alpha
			\left<
				S_{l+1} - S_l
			\right>_\alpha
			.
	    \end{align}
	\item Calculate the entropic $C$-functions \eqref{eq:EntropicCFunctionDef} by
		\begin{align}\label{eqn:EntropicClattice}
			C(l-\nicefrac{a}{2}) &= \frac{(l-\nicefrac{a}{2})^{3}}{aL^2}
			\frac{\partial S^{(2)}}{\partial l}\bigg\vert_{l-\nicefrac{a}{2}}
			.
		\end{align}
\end{enumerate}

\section{Numerical results} \label{sec:Results}
We have shown in the previous section how to calculate R\'enyi entropies on the lattice.
In this section we will first show the details of our lattice calculation and afterwards we are going to present results for the entropic $C$-function in $SU(2)$, $SU(3)$ and $SU(4)$ pure Yang-Mills theory.

To implement the method described in the last section, we use a standard Wilson gauge action given by
\begin{align}
	S[U] = \frac{\beta}{N_c}\sum_{U_{\mu\nu}} \text{Re }\tr\left(1 - U_{\mu\nu}\right)
	\eqnPunct .
\end{align}
We are using a pseudo-heatbath algorithm to update the gauge configurations. Doing this we have to calculate the closing plaquettes for the link which we want to update. It is crucial in this step that we carefully choose the nearest neighbors of the lattice sites due to the non-trivial lattice topology. 

To generate random $SU(N_c)$ matrices we use the Cabbibo-Marinari algorithm~\cite{Cabbibo:1982aaa}.
To avoid autocorrelation we wait 100 sweeps between successive measurements.
For all sets of parameters we use a lattice of size
$
	N_s^3 \times \left( q \times N_t \right)
	= 
	16^3 \times \left( 2 \times 16 \right)
$.
Since the resolution of the entropic $C$-function in $l$ would be very poor if we only considered one single $\beta$ value, we have to set the scale for each lattice and compare results for several different lattice spacings. To do so we use the string tension $\sqrt{\sigma}$ as determined in~\cite{Lucini:2005vg} for a range of $\beta$ values and then measure everything in units of $\sqrt{\sigma}$. 

The parameters of the gauge field configurations which we have used in our analysis are summarized in table~\ref{table:Configurations}. The numbers in these tables are the numbers of configurations we gathered for each $\alpha$ value.\\
\begingroup
\squeezetable
\begin{table*}[htbp]
		\input{tableSUN}
		\caption{
			$SU(2)$, $SU(3)$ and $SU(4)$ configurations for the used $\beta$ values and cut lengths $l$.
			The numbers are the configurations simulated for each value of $\alpha$. To get the total number of configurations one has to multiply the $SU(2)$ and $SU(3)$ configurations by $11$ and the $SU(4)$ ones by $21$.
		}
		\label{table:Configurations}
\end{table*}
\endgroup%
As mentioned in the last section (cf. equation~\eref{eqn:EntropicClattice}), we have to integrate the interpolating action over $\alpha$. Therefore, we have to study carefully how many interpolating points we have to consider. We did this by increasing the number of points in $\alpha$ and then checking the integral value for convergence. As in~\cite{Itou:2015cyu} we find that we need $11$ interpolating points for $SU(3)$, but we need $21$ for $SU(4)$. For $SU(2)$ we also use $11$ interpolating points instead of the $6$ which were used in~\cite{Buividovich:2008kq}. The increased number of $\alpha$ values can be explained by the larger lattice volume and the, therefore, larger integrand.
Besides the increased computational cost due to larger color matrices, this gives an additional factor of $\sim 2$ for $N_c > 2$ in computation time.

\begin{figure*}[htbp]
        \includegraphics[width=0.8\linewidth]{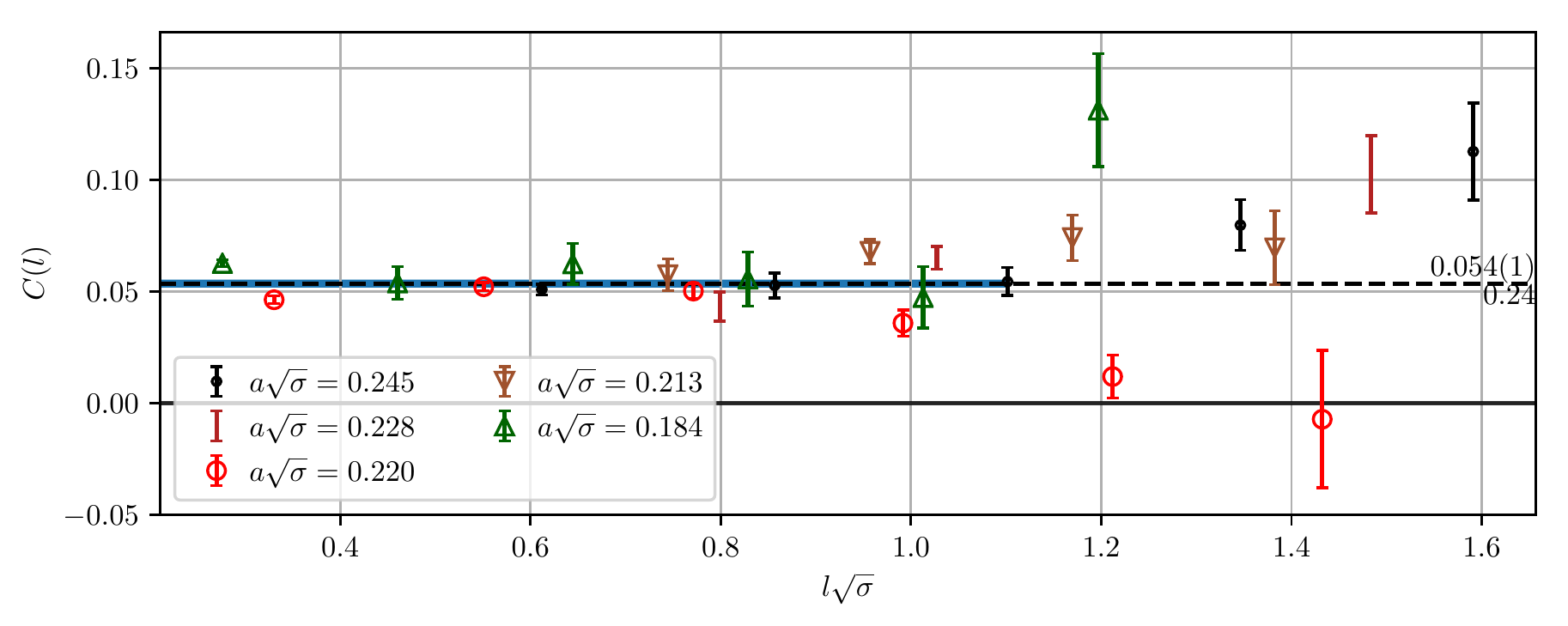}\\
		\includegraphics[width=0.8\linewidth]{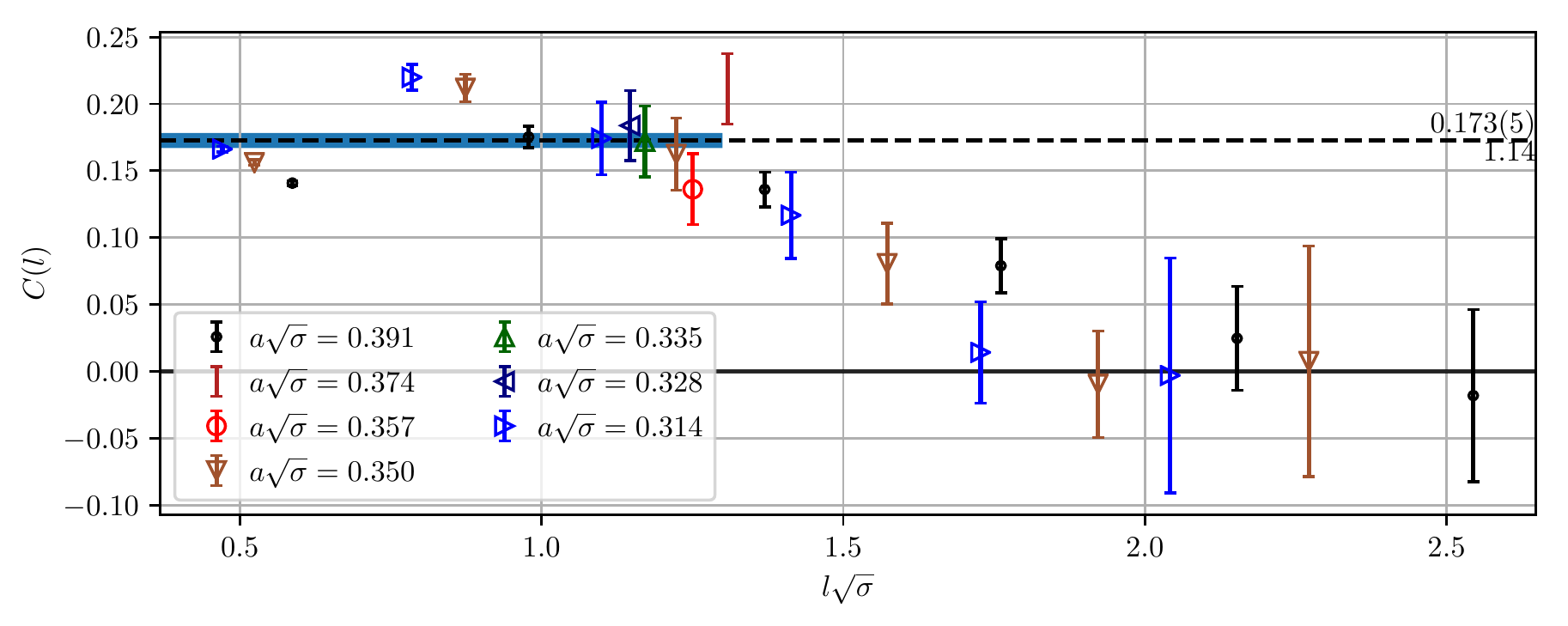}\\
		\includegraphics[width=0.8\linewidth]{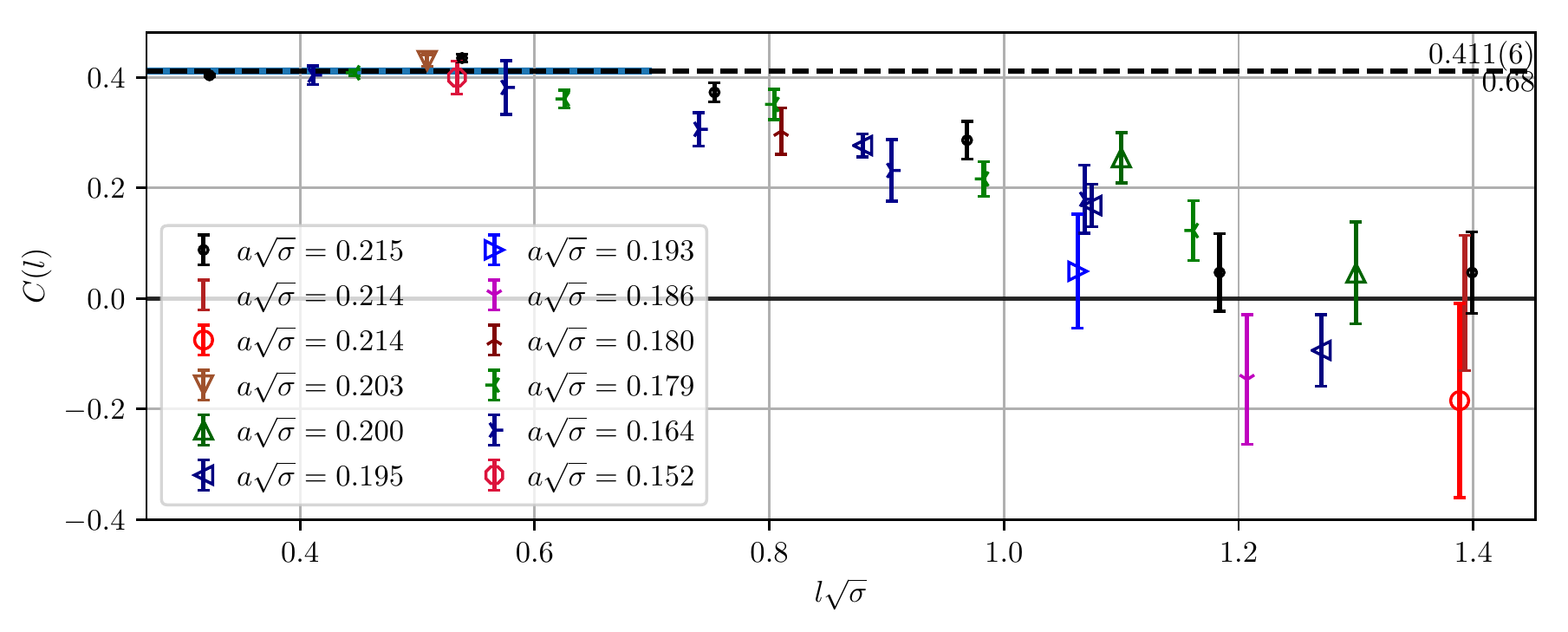}
	\caption{
		Entropic $C$-function for $SU(2)$, $SU(3)$ and $SU(4)$ (from top to bottom) lattice gauge theories. The black dashed line with the blue error band is a fit to a constant for the first data points where $l\sqrt{\sigma}$ is less than $\lmaxTwo$, $\lmaxThree$ and $\lmaxFour$ for $SU(2)$, $SU(3)$ and $SU(4)$ respectively. The numerical value with error is given on the very right of each plot. The value below gives the $\nicefrac{\chi^2}{d.o.f.}$ for this fit.
	}
	\label{fig:EntropicC}
\end{figure*}

The entropic $C$-function which we obtain from our simulations are shown in figure~\ref{fig:EntropicC}. In these plots, the different colors and markers correspond to different lattice spacings. 

For sufficiently small values of $l$, the $C$-function appears to be approximately constant for all values of $N_c$. Since the entropic $C$-function (\ref{eq:EntropicCFunctionDef}) is defined in such a way that it yields a constant for non-interacting conformally invariant fields, this behavior could be expected in the short-distance asymptotic freedom regime where entropy is saturated by weakly interacting massless gluons. To extract the corresponding constant values of the $C$-function, we fitted a constant function to the first data points where $l\sqrt{\sigma}$ is less than $\lmaxTwo$, $\lmaxThree$ and $\lmaxFour$ for $SU(2)$, $SU(3)$ and $SU(4)$ respectively. We also varied the fitting range slightly and found a mild dependence. This fit is shown by the dashed black line with the blue error band. The constant values and $\nicefrac{\chi^2}{d.o.f.}$ for these fits are given above and below this line, respectively.

At larger cut lengths $l$ and for $N_c = 3, 4$, we observe the expected monotonic (within statistical errors) decrease of $C(l)$ with $l$ down to a value which is compatible with zero within our statistical errors. This fall-off does not look like the discontinuous transition predicted by a holographic calculation \cite{Klebanov:2007ws}, but this is expectable for a finite lattice and finite $N_c$. 

\begin{figure}[htbp]
	\includegraphics[width=\linewidth]{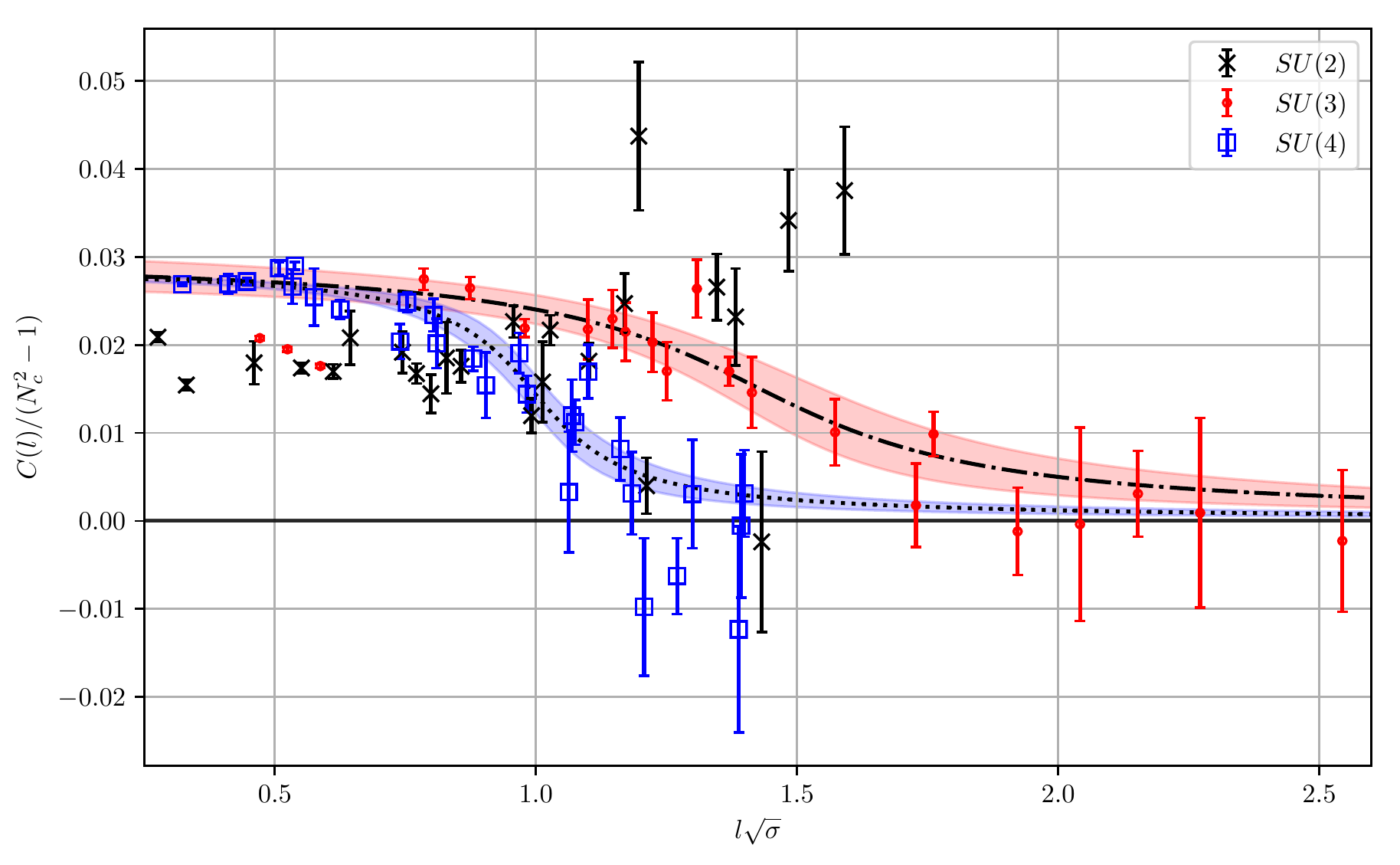}
	\caption{
		The entropic $C$-function \eqref{SEE_scaling} for $SU(2)$, $SU(3)$ and $SU(4)$ rescaled by $N_c^2-1$, the number of gluons. The fits to these data are according to a regularized step function~\eref{eqn:RegStep}. The first three red points ($SU(3)$) were excluded from the fit due to large discretization errors, but show a trend towards the fit in the continuum limit. 
	}
	\label{fig:Compare234}
\end{figure}

On the other hand, for $SU(2)$ the data looks qualitatively different and does not exhibit any clear trend for $l \sqrt{\sigma} \gtrsim 1$. It is striking however that for $a \sqrt{\sigma}=0.220$, the data points show the expected transition around $l\sqrt{\sigma}=1.0$. One possible explanation of the absence of a clear trend in the data for other lattice spacings is that $a \sqrt{\sigma}\approx 0.220$ gives only a small window of lattice spacings where both finite volume and discretization errors are small enough. For finite volume effects, this is supported by the temperature on our lattice being already half the critical temperature for $a \sqrt{\sigma}\approx 0.18$ \cite{Fingberg:1992ju}, see also \cite{Giudice:2017dor}. We emphasize that this discussion is only qualitative so far and needs to be revisited using larger lattices, which is however outside of the scope of the current paper. In the concluding section \ref{sec:Conclusion} we also discuss another possible explanation related to the different orders of finite-temperature phase transitions in $SU(2)$ and $SU(N_c > 2)$ gauge theories.
To compare the data points for different gauge groups, we jointly plot in figure~\ref{fig:Compare234} the data for the entropic $C$-function \eqref{SEE_scaling} for $SU(2)$, $SU(3)$ and $SU(4)$. For a better comparison across different gauge groups, we rescale the data by a factor $N_c^2 - 1$, which is the expected scaling in the small-cut regime. In order to keep the plot readable, we no longer distinguish data points with different values of lattice spacing. For small cut lengths, this plot again confirms the scaling of entanglement entropy with $N_c^2 - 1$.
\\

In order to check whether the transition between small-cut and large-cut regimes becomes steeper for larger $N_c$, as predicted by the holographic calculation \cite{Klebanov:2007ws}, we fit the data to a regularized step function given by
\begin{align}\label{eqn:RegStep}
	A
	\left[
		\frac{1}{2} + \frac{1}{\pi}\arctan\left(\frac{x-C}{B}\right)
	\right]
	\eqnPunct .
\end{align}
The resulting fit with error bands is visualized in figure~\ref{fig:Compare234} by the black lines and the fit parameters are given by
\begin{center}
	\begin{tabular}{| c | c | c | c |}\hline
		$N_c$ & $A$ & $B$ & $C$ \\\hline\hline
		$3$ & $\AThree $ & $\BThree$ & $\CThree$\\\hline
		$4$ & $\AFour $ & $\BFour$ & $\CFour$\\\hline
	\end{tabular}
\end{center}
From these results we can calculate the maximal slope $m_\text{max} = \frac{A}{B\pi}$ which gives
\begin{align}
        m_\text{max}^{SU(3)} &= \SlopeThree \label{eq:MaxSlope3} \\
        m_\text{max}^{SU(4)} &= \SlopeFour \label{eq:MaxSlope4}
        \eqnPunct .
\end{align}
    
While the statistics is not large enough to reach a definite conclusion, we observe a trend towards a steeper slope at $N_c = 4$, which is consistent with the scenario of discontinuous transition in the large $N_c$ limit.

\section{Discussion} \label{sec:Conclusion}

In this paper, we have calculated the second R\'enyi entropy $S^{(2)}$ as function of the width $l$ of the slab-like entangled region in pure $SU(N_c)$ Yang-Mills theory for $N_c = 2, 3,4$ using lattice techniques. $S^{(2)}$ here serves as an approximation to the entanglement entropy, which is hard to compute directly.

We have confirmed that at short distances the UV-finite part of the entanglement entropy scales proportional to the number of gluon states, $N_c^2 - 1$. At larger values of $l$ the entropy drops significantly, in agreement with the absence of colorful states in the low-energy confinement regime. We found that the decrease of the R\'enyi entropy at larger $l$ is steeper for $N_c = 4$ than for $N_c = 3$, which might be an indication that in the large-$N_c$ limit the transition between small-$l$ and large-$l$ regimes is indeed discontinuous, as predicted by holographic models \cite{Klebanov:2007ws}. 

Let us note that since for finite $N_c$ phase transitions cannot happen in a gauge theory on a finite lattice, we could not expect a discontinuous transition in our data. On the other hand, since for sufficiently large lattices the large-$N_c$ limit mimics the infinite-volume limit \cite{Eguchi:82:1,Narayanan:hep-lat/0303023}, our findings can imply two different scenarios. One possibility is that the entanglement entropy is a discontinuous function of $l$ for any $N_c$, but the discontinuity can only be seen by working on sufficiently large lattices and taking the proper infinite-volume limit. Another possibility is that the discontinuity appears only for infinite $N_c$, since the predictions of \cite{Klebanov:2007ws} based on the AdS/CFT correspondence and a Hagedorn-type spectrum are strictly speaking only valid in this limit. In this case, a careful study of even larger values of $N_c$ is required. Needless to say, one also cannot exclude that the predictions of~\cite{Klebanov:2007ws} are not directly applicable to pure gauge theories, but rather require supersymmetry.

An interesting observation is that the data for $SU(2)$ gauge theory appear to be qualitatively different from that for the $SU(3)$ and $SU(4)$ in that the approximate entropic $C$-function extracted from the second R\'enyi entropy does not decrease to zero at large cut lengths, but rather exhibits a fluctuating behavior without any clear trend. While this difference might well originate in lattice artifacts, it is also possible that a different behavior of entanglement entropy in $SU(2)$ and $SU(N_c > 2)$ gauge theories is a physical feature related to the different orders of finite-temperature deconfinement phase transitions in these theories. It is not unreasonable to conjecture that the behavior of entanglement entropy near critical cut length might be related to the behavior of thermal entropy in the vicinity of a finite-temperature phase transition, as both transitions appear to be qualitatively similar for theories with a Hagedorn-type spectrum of states \cite{Klebanov:2007ws}. In the vicinity of a second-order thermal phase transition in $SU(2)$ gauge theory one can expect an enhancement of entanglement entropy and its fluctuations due to the divergence of correlation length \cite{Anber:2018ohz}, which is absent for first-order finite-temperature transitions in $SU(N_c > 2)$ gauge theories.

For future work, it would be of obvious interest to increase the statistics and perform a continuum limit. Going to larger $N_c$ may reveal a steeper slope and thus a more abrupt transition, which our data already hints at. However, the computational cost to accurately evaluate the integral \eqref{eqn:DifferenceFreeEnergies} increases not only due to the obvious $N_c$-dependence, but also due to the need for a finer lattice of $\alpha$-values at larger $N_c$. For $N_c = 3,4$, this so far resulted in an additional factor of $2^{N_c}$. This observation makes it questionable whether accurate data at larger $N_c$ will be available in the near future. 

On the other hand, improving our understanding of the behavior of the entropic $C$-function for $SU(2)$ Yang-Mills theory is in reach using moderate computational resources and certainly interesting to pursue. 

Also the holographic calculations of the entanglement entropy can be brought closer to the lattice setup by considering corrections due to finite 't Hooft coupling. Instead of entanglement entropy, one can also calculate the R\'enyi entropy within holographic models \cite{Dong:2016fnf}, which might exhibit a behavior closer to what we have found numerically.

	\acknowledgments
	N.~Bodendorfer was supported by an International Junior Research Group grant of the Elite Network of Bavaria. P.~Buividovich was supported by the Heisenberg Fellowship from the German Research Foundation, project BU2626/3-1. A. Rabenstein and A. Sch\"afer were supported by SFB/TRR-55 ``Hadron Physics from Lattice QCD''. We thank the ECT* for its hospitality during the workshop ``Quantum Gravity meets Lattice QFT'' where part of this work was done. The numerical simulations were performed on ATHENE the HPC cluster of the Regensburg University Compute Centre. We thank Stefano Piemonte and Mohamed Anber for valuable discussions regarding $SU(2)$ Yang-Mills theory.

\end{document}

%% file: tikzBipartition.tex
\begin{figure}
\centering
\begin{tikzpicture}
	\def\shift{0.2}
	\newif\ifinlist
	\def\entangled{0/0,0/1,1/0,1/1,2/0}
	\tikzset{
		continuum/.pic={
			\draw[fill=none,thick,rounded corners](-1.5,-1.5) -- (-1.5,3.5) -- (3.5,3.5) -- (3.5,-1.5) -- cycle ;
		},
		links/.pic={
			\foreach \x in {-1,0,...,2}
			{

				\foreach \y in {-1,0,...,2}
				{
					\draw[->] ($(\x,\y)+(0.1,0)$) -- ($(\x,\y)+(0.9,0)$);
					\draw[->] ($(\x,\y)+(0,0.1)$) -- ($(\x,\y)+(0,0.9)$);
				}
			}
		},
		boundary/.pic={
			\draw[rounded corners=1ex]
				(0-\shift,0-\shift)
				--
				(2+\shift,0-\shift)
				--
				(2+\shift,1+\shift)
				--
				(1+\shift,1+\shift)
				--
				(1+\shift,2-\shift)
				--
				(0-\shift,2-\shift)
				--
				cycle
				;
		},
		links/.pic={
			\foreach \x in {-1,0,...,2}
			{

				\foreach \y in {-1,0,...,2}
				{
					\draw[->] ($(\x,\y)+(0.1,0)$) -- ($(\x,\y)+(0.9,0)$);
					\draw[->] ($(\x,\y)+(0,0.1)$) -- ($(\x,\y)+(0,0.9)$);
				}
			}
		},
		latticeIN/.pic={
			\foreach \x/\y in \entangled
			{
				\draw[fill=black] (\x,\y) circle(2pt);
			}
		},
		boundaryLink/.pic={
			\draw[rounded corners=0.7ex] (0,0) -- (-0 , 2)
				-- (1,2)
				-- ++(0,-1)
				-- ++(1,0)
				-- ++(0,-1)
				-- cycle
				;
		},
		latticeOUT/.pic={
			\foreach \x in {-1,0,...,3}
			{

				\foreach \y in {-1,0,...,3}
				{
					\inlisttrue
					\foreach \Xcompare/\Ycompare in \entangled 
					{
						\ifthenelse{\x = \Xcompare \AND \y = \Ycompare }{\global\inlistfalse};
					}
					\ifinlist
					\draw[fill=black] (\x,\y) circle(2pt);
					\fi
				}
			}
		},
	}
	\def\scale{0.6}
	\pic[scale=\scale] at(0,0) {continuum};
	\pic[scale=\scale] at(0,0) {boundary};
	\node  at ($\scale*(-1,3)$) {$\Sigma_t$};
	\node[above right] at ($\scale*(2.5,2.5)$) {$\bar{\mathcal{A}}$};
	\node[] at ($\scale*(1,0.6)$) {$\mathcal{A}$};
	\node[] at ($\scale*(2.6,0.5)$) {$\partial\mathcal{A}$};
	\pic[scale=\scale] at($\scale*(7,0)$) {continuum};
	\pic[scale=\scale] at($\scale*(7,0)$) {latticeIN};
	\pic[scale=\scale] at($\scale*(7,0)$) {latticeOUT};
	\pic[scale=\scale] at($\scale*(7,0)$) {links};
	\pic[scale=\scale,line width = 1.4pt,red] at($\scale*(7,0)$) {boundaryLink};
	\pic[scale=\scale,line width = 0.7pt,dashed,red] at ($\scale*(7,0)$) {boundary};

\end{tikzpicture}
\caption{
	The left figure shows the bipartition of a fixed time slice $\Sigma_t$ in regions $\mathcal{A}$ and $\bar{\mathcal{A}}$.
	The boundary between the two regions $\partial\mathcal{A}$ is called entangling surface. The right figure shows the maximally gauge invariant way of how to factorize the Hilbert space for a gauge theory~\cite{Aoki:2015bsa}.
	The bipartition is chosen such that we cut along the links (solid line), because other possibilites (dashed line) would cut links and thus violate gauge invariance.
}
\label{fig:Bipartition}
\end{figure}

%% file: tikz_density.tex
\foreach \x in {-1,0,1,2}
{
	\foreach \y in {-1,0,1,2}
	{
		\draw[fill=black] (\x,\y) circle(2pt);
	}
	\draw (\x,-1) -- (\x,2);
	\draw (-1,\x) -- (2,\x);
}
\draw[rounded corners = 4pt,dashed] (-1.2,1.8) 
	--++ (0,0.4)
	--++ (3.4,0)
	--++ (0,-0.4)
	-- cycle;
\draw[rounded corners = 4pt,dashed] (-1.2,-1.2) 
	--++ (0,0.4)
	--++ (3.4,0)
	--++ (0,-0.4)
	-- cycle;
\node[right] at (2.2,-1) {$\ket{\psi_1}$};
\node[right] at (2.2,2) {$\ket{\psi_2}$};

%% file: tikz_density_reduced.tex
\foreach \x in {-1,0,1,2}
{
	\foreach \y in {-1,0,1,2}
	{
		\draw[fill=black] (\x,\y) circle(2pt);
	}
	\draw (\x,-1) -- (\x,2);
	\draw (-1,\x) -- (2,\x);
}

\draw[rounded corners = 4pt,dashed] (1,1.8) 
	--++ (0,0.4)
	--++ (1.2,0)
	--++ (0,-0.4)
	-- cycle;
\draw[rounded corners = 4pt,dashed] (1,-1.2) 
	--++ (0,0.4)
	--++ (1.2,0)
	--++ (0,-0.4)
	-- cycle;
\draw[rounded corners = 4pt,dashed] (-1.2,1.8) 
	--++ (0,0.4)
	--++ (2.2,0)
	--++ (0,-0.4)
	-- cycle;
\draw[rounded corners = 4pt,dashed] (-1.2,-1.2) 
	--++ (0,0.4)
	--++ (2.2,0)
	--++ (0,-0.4)
	-- cycle;
\node[right] at (2.2,-1) {$\ket{\psi_1}$};
\node[right] at (2.2,2) {$\ket{\psi_2}$};
\node[left] at (-1.2,-1) {$\ket{\psi}$};
\node[left] at (-1.2,2) {$\ket{\psi}$};

%% file: tikz_density_reduced_deformed.tex
\foreach \x in {-1,0,1}
{
\foreach \y in {-1,0,1,2}
{
\draw[fill=black] (\y,\x) circle(2pt);
}
\draw (-1,\x) -- (2,\x);
\draw (\x,-1) -- (\x,1);
\draw[fill=black] (1,2) circle(2pt);
\draw[fill=black] (2,2) circle(2pt);
\draw (1,2) -- (2,2);
};
\draw (1,1) -- (1,2);
\draw (2,-1) -- (2,2);
\draw[rounded corners = 4pt,dashed] (1,1.8) 
--++ (0,0.4)
--++ (1.2,0)
--++ (0,-0.4)
-- cycle;
\draw[rounded corners = 4pt,dashed] (1,-1.2) 
--++ (0,0.4)
--++ (1.2,0)
--++ (0,-0.4)
-- cycle;
\node[right] at (2.2,-1) {$\ket{\psi_1}$};
\node[right] at (2.2,2) {$\ket{\psi_2}$};
\tikzClose{(-1,1)}{(-1,-1)};
\tikzClose{(0,1)}{(0,-1)};
\draw (0,-1) -- (1,2);
\node[rotate=-45] at ($(0,-1)+(0.3,0.2)$) {$=$};

%% file: tikz_tr_density_reduced_squared.tex
\tikzset{
	lattice/.pic = {
		\foreach \x in {-1,0,1}
		{
			\foreach \y in {-1,0,1,2}
			{
				\draw[fill=black] (\y,\x) circle(2pt);
			}
			\draw (-1,\x) -- (2,\x);
			\draw (\x,-1) -- (\x,1);
		};
		\tikzClose{(-1,1)}{(-1,-1)};
		\tikzClose{(0,1)}{(0,-1)};
		\node[rotate=-45] at ($(0,-1)+(0.3,0.2)$) {$=$};
	}
}
\pic[scale=0.7] at (0,0) {lattice};
\pic[scale=0.7] at (0,3) {lattice};
\draw (2,-1) -- (2,4);
\draw (1,-1) -- (1,4);

\tikzClose{(1,4)}{(1,-1)};
\tikzClose{(2,4)}{(2,-1)};

\draw (0,-1) -- (1,2);
\draw (0,2) -- ($(0,2) ! 0.73 ! (1,5)$);
\draw (1,-1) -- ($(1,-1) ! 0.15 ! (0,-4)$);

%% file: tikz_periodicity.tex
\def\shift{.3};
\foreach \y in {0,...,6}
{
	\foreach \x in {0,...,6}
	{
		\draw[fill=black] (\x+\y*\shift,\y) circle(2pt);
	}
	\draw (0+\y*\shift,\y) -- ( \y*\shift + 6,\y);
}
\foreach \x in {0,...,6}
{
	\draw (\x,0) -- (\x+6*\shift,6);
}
\draw [fill=fakultaet,opacity=0.3] (0,0) -- (3,0) -- (3+3*\shift,3) -- (0+3*\shift,3) -- cycle;
\draw [fill=fakultaet,opacity=0.3] (0+3*\shift,3) -- (3+3*\shift,3) -- (3+6*\shift,6) -- (0+6*\shift,6) -- cycle;
\draw [fill=red,opacity=0.3] (3+0*\shift,0) -- (6+0*\shift,0) -- (6+6*\shift,6) -- (3+6*\shift,6) -- cycle;

\draw [thick,red] (0+3*\shift,3) -- (3+3*\shift,3);
\foreach \x in {0,1,2,3}
{
	\draw[fill=black] (\x+3*\shift,3) circle(2pt);
}
\foreach \x in {0,1,2}
{
	\draw --plot [only marks,mark=x, mark options={color=red},mark size=6pt] coordinates {(3+ \x + 0.5 + 0*\shift,0)};
	\draw --plot [only marks,mark=x, mark options={color=red},mark size=6pt] coordinates {(3+ \x + 0.5 + 6*\shift,6)};
	\draw --plot [only marks,mark=square, mark options={color=fakultaet},mark size=4pt] coordinates {(0+ \x + 0.5 + 0*\shift,0)};
	\draw --plot [only marks,mark=square, mark options={color=fakultaet},mark size=4pt] coordinates {(0+ \x + 0.5 + 2.9*\shift,2.9)};
	\draw --plot [only marks,mark=triangle, mark options={color=black},mark size=5pt] coordinates {(0+ \x + 0.5 + 3.1*\shift,3.1)};
	\draw --plot [only marks,mark=triangle, mark options={color=black},mark size=5pt] coordinates {(0+ \x + 0.5 + 6*\shift,6)};
}
\draw [<->] (6.5 + 0*\shift,0) -- (6.5+6*\shift,6);
\node[right] at ($(6.5 + 0*\shift,0) !0.5! (6.5+6*\shift,6)$) {\small $2T$};
\draw [<->] (-0.5 + 0*\shift,0) -- (-0.5+3*\shift,3);
\node[left] at ($(-0.5 + 0*\shift,0) !0.5! (-0.5+3*\shift,3)$) {\small $T$};
\draw [<->] (-0.5 + 3*\shift,3) -- (-0.5+6*\shift,6);
\node[left] at ($(-0.5 + 3*\shift,3) !0.5! (-0.5+6*\shift,6)$) {\small $T$};

\def\ypos{-1}
\draw[<->,thick] (0+\ypos*\shift,\ypos) -- (3+\ypos*\shift,\ypos);
\node[below] at ($(0+\ypos*\shift,\ypos) !0.5! (3+\ypos*\shift,\ypos)$) {\small$\bar{\mathcal{A}}$};
\draw[<->,thick] (3+\ypos*\shift,\ypos) -- (6+\ypos*\shift,\ypos);
\node[below] at ($(3+\ypos*\shift,\ypos) !0.5! (6+\ypos*\shift,\ypos)$) {\small${\mathcal{A}}$};

%% file: tableSUN.tex
\begin{tabular}{ | c  | c ||  c |  c |  c |  c |  c |  c | }
	\multicolumn{8}{c}{\vspace{1em}\textbf{Configurations for $\mathbf{SU(2)}$}}\\
	\hline
	$\beta$
	&\diagbox[innerwidth=1.5cm]{$a\sqrt{\sigma}$}{$l$}
	& $2$ & $3$ & $4$ & $5$ & $6$ & $7$\\\hline\hline
    $2.420$ & $ 0.245 $ & $ $ & $414,720$ & $414,720$ & $1,708,280$ & $1,437,288$ &     $1,432,656$\\\hline
    $2.440$ & $ 0.228 $ & $ $ & $ $ & $276,480$ & $1,928,208$ & $1,808,640$ & $1,763,604$\\\hline
    $2.450$ & $ 0.220 $ & $172,800$ & $1,013,760$ & $1,642,921$ & $2,582,780$ & $2,583,793$ &     $2,561,809$\\\hline
    $2.460$ & $ 0.213 $ & $ $ & $ $ & $276,480$ & $1,840,128$ & $1,826,752$ & $1,750,272$\\\hline
    $2.500$ & $ 0.184 $ & $172,800$ & $172,800$ & $449,280$ & $829,440$ & $1,720,584$ &     $1,723,629$\\\hline
   \multicolumn{8}{c}{}\\
	\multicolumn{8}{c}{\vspace{1em}\textbf{Configurations for $\mathbf{SU(3)}$}}\\
	\hline
	$\beta$
    &\diagbox[innerwidth=1.5cm]{$a\sqrt{\sigma}$}{$l$}
	& $2$ & $3$ & $4$ & $5$ & $6$ & $7$\\\hline\hline
	$5.700$ & 0.391&$44,976$ & $44,976$ & $130,656$ & $253,056$ & $218,328$ & $230,568$\\\hline
	$5.720$ & 0.374&$ $ & $ $ & $34,728$ & $ $ & $ $ & $ $\\\hline
	$5.740$ & 0.357&$ $ & $ $ & $34,728$ & $ $ & $ $ & $ $\\\hline
	$5.750$ & 0.350&$34,728$ & $34,728$ & $34,728$ & $120,408$ & $208,080$ & $134,640$\\\hline
	$5.770$ & 0.335&$ $ & $ $ & $34,728$ & $ $ & $ $ & $ $\\\hline
	$5.780$ & 0.328&$ $ & $ $ & $34,728$ & $ $ & $ $ & $ $\\\hline
	$5.800$ & 0.314&$34,728$ & $34,728$ & $34,728$ & $120,408$ & $255,048$ & $134,640$\\\hline
	\multicolumn{8}{c}{}\\
	\multicolumn{8}{c}{\vspace{1em}\textbf{Configurations for $\mathbf{SU(4)}$}}\\
	\hline
	$\beta$
	&\diagbox[innerwidth=1.5cm]{$a\sqrt{\sigma}$}{$l$}
	& $2$ & $3$ & $4$ & $5$ & $6$ & $7$\\\hline\hline
	$11.000$ & 0.215&$24,480$ & $70,560$ & $93,600$ & $93,600$ & $69,120$ & $184,320$\\\hline
	$11.004$ & 0.214&$ $ & $ $ & $ $ & $ $ & $ $ & $57,600$\\\hline
	$11.008$ & 0.214&$ $ & $ $ & $ $ & $ $ & $ $ & $30,744$\\\hline
	$11.058$ & 0.203&$ $ & $26,650$ & $ $ & $ $ & $ $ & $ $\\\hline
	$11.075$ & 0.200&$ $ & $ $ & $ $ & $ $ & $160,221$ & $115,200$\\\hline
	$11.100$ & 0.195&$ $ & $ $ & $ $ & $253,440$ & $253,440$ & $253,440$\\\hline
	$11.112$ & 0.193&$ $ & $ $ & $ $ & $ $ & $30,744$ & $ $\\\hline
	$11.156$ & 0.186&$ $ & $ $ & $ $ & $ $ & $ $ & $57,600$\\\hline
	$11.192$ & 0.180&$ $ & $ $ & $ $ & $33,961$ & $ $ & $ $\\\hline
	$11.200$ & 0.179&$ $ & $103,680$ & $103,680$ & $149,760$ & $357,120$ & $357,120$\\\hline
	$11.300$ & 0.164&$ $ & $11,520$ & $11,520$ & $126,720$ & $126,720$ & $264,960$\\\hline
	$11.398$ & 0.152&$ $ & $ $ & $31,704$ & $ $ & $ $ & $ $\\\hline
\end{tabular}